\documentclass[a4paper,11pt]{article}
\usepackage[utf8]{inputenc}
\usepackage{geometry}
\usepackage{authblk}
\usepackage{cite}
\usepackage{setspace}
\usepackage{amsmath}
\usepackage{graphicx}
\usepackage{hyperref}
\usepackage{caption}
\usepackage{subcaption}

\usepackage{tabularx} 
\usepackage{xcolor}   
\usepackage{amsmath}  
\usepackage{graphicx} 
\usepackage{caption}
\usepackage{subcaption}

\title{\vspace{-2cm}\textbf{From Safeguards Application to Fundamental Physics: Advancements in Reactor Neutrino Detection with the $\nu$-Angra Experiment}}

\author[1]{E. Kemp}
\author[2]{M. P. Albuquerque}
\author[2]{J. C. Anjos}
\author[3]{P. Chimenti}
\author[4]{L. F. G. Gonzalez}
\author[5]{G. P. Guedes}
\author[6,7]{P. V. Guillaumon}
\author[2,8]{H. P. Lima Jr.}
\author[2]{A. Massafferri}
\author[2,9]{L. M. Domingues Mendes}
\author[10]{R. A. Nóbrega}
\author[11]{I. M. Pepe}
\author[1]{W. V. Santos}

\affil[1]{\footnotesize Instituto de Física Gleb Wataghin, Universidade Estadual de Campinas, Campinas, 13083-859, SP, Brazil}
\affil[2]{\footnotesize Centro Brasileiro de Pesquisas Físicas, Rio de Janeiro, 22290-180, RJ, Brazil}
\affil[3]{\footnotesize Departamento de Física, Universidade Estadual de Londrina, Londrina, 86057-970, PR, Brazil}
\affil[4]{\footnotesize Instituto de Computação, Universidade Estadual de Campinas, Campinas, 13083-852, SP, Brazil}
\affil[5]{\footnotesize Universidade Estadual de Feira de Santana, Feira de Santana, 44036-900, BA, Brazil}
\affil[6]{\footnotesize Instituto de Física, Universidade de São Paulo, São Paulo, 05508-090, SP, Brazil}
\affil[7]{\footnotesize Max-Planck-Institute für Physik, München, 80805, Germany}
\affil[8]{\footnotesize Astroparticle Physics Division, Gran Sasso Science Institute, L’Aquila, 67100, Italy}
\affil[9]{\footnotesize Laboratório de Instrumentação e Física Experimental de Partículas, Lisboa, 1649-003, Portugal}
\affil[10]{\footnotesize Departamento de Circuitos Elétricos, Universidade Federal de Juiz de Fora, Juiz de Fora, 36036-900, MG, Brazil}
\affil[11]{\footnotesize Instituto de Física, Universidade Federal da Bahia, Salvador, 40170-115, BA, Brazil}

\date{\vspace{-1cm}July 26, 2024}

\begin{document}

\maketitle

\begin{abstract}
  Operating on a surface with high noise rates and requiring susceptible, yet small-scale detectors, the Neutrinos-Angra detector is an excellent platform for technological development and expertise in new detection methods. This report details the primary features of the detector, the electronics involved, and preliminary physics results from the operational phase, particularly the ON-OFF analysis comparing the signals with the reactor in operation (ON) and during the maintenance shut-down (OFF), demonstrating the detector’s capability to monitor reactor activity. Additionally, we will briefly discuss the prospects of using a cryogenic calorimeter to detect neutrinos via Coherent Elastic Neutrino-Nucleus Scattering (CEvNS), highlighting potential advancements in neutrino detection technology. Looking ahead, the project promises to play a crucial role in the integration of Latin American scientists and engineers into global scientific collaborations, significantly contributing to the LASF4RI and the broader HECAP strategic framework.
\end{abstract}

\textbf{\Large Thematic Areas}: Neutrino Physics, Instrumentation and Computing.

\section*{Contact Person}
\begin{itemize}
  \item Name: Ernesto Kemp
  \item Email: \texttt{kemp@unicamp.br}
\end{itemize}

\section{Scientific context}

Nuclear reactors have been fundamental in experimental neutrino physics because of their role as prolific sources of man-made neutrinos. The confirmation of the neutrino hypothesis in 1956 \cite{Cowan1956}, along with subsequent experiments, has significantly deepened our understanding of neutrino oscillation parameters \cite{Eguchi2003, An2012, Abe2012, Ahn2012}.

Neutrinos emitted from nuclear reactors are directly related to the fission of heavy nuclei: each fission event releases a known fraction of energy and is followed by neutrino emission. By monitoring this neutrino flux, one can estimate the fission rate and the reactor's thermal power, making the neutrino flux a direct measure of reactor power. The idea of using neutrinos for remote monitoring of reactor thermal power emerged in the mid-1970s \cite{Mikaelian1978, Borovoi1978}. Early demonstration experiments, such as those at the Rovno nuclear power plant in Ukraine, clearly showed the correlation between neutrino count rates and reactor activity \cite{Korovkin1984, Korovkin1988}, paving the way for neutrino-based monitoring. This capability makes neutrinos reliable probes for physical processes in reactors, aligning with the International Atomic Energy Agency (IAEA) non-proliferation safeguards. Neutrino detectors enable remote monitoring of reactor activities, avoiding the need for intrusive operations in restricted areas of nuclear plants. Summaries of global efforts in this area can be found in various references \cite{Bowden2008, Carr2018}.

Initially considered for studying neutrino oscillations \cite{Anjos2006}, the Angra dos Reis nuclear plant became a site for the $\nu$-Angra experiment focused on non-proliferation safeguards. The Angra-II reactor generates a steady-state neutrino flux of approximately $1.21 \times 10^{20}$ s$^{-1}$ \cite{Angra2006}, useful for monitoring reactor activity and estimating the core's thermal power.

The $\nu$-Angra experiment aims to develop cost-effective technology for monitoring reactor power and neutrino spectral evolution to detect plutonium fractions, critical for diversion detection. Stable data acquisition is crucial, adhering to IAEA safeguards \cite{IAEATech2009}. The electronics have been validated for stable, long-term operation.

A challenge for neutrino experiments, reactor-based or not, is background radiation, mainly neutrons and cosmic muons, mimicking neutrino signals. The typical solution is installing detectors in large underground caverns, using rock and soil as shields. Studies show significant reductions in vertical muon intensity at sufficient depths \cite{Grieder2001}.
The $\nu$-Angra detector is installed on the surface per plant safety rules, challenging its signal-to-noise ratio for effective reactor monitoring.

The detector became operational in late 2018 \cite{Lima2019}. This document outlines the detector's characteristics, initial data analysis, and future plans, such as enhancements to the existing detector and the potential to run a cryogenic calorimeter with quantum sensor readout to maintain safeguard purposes and to investigate the Coherent Elastic Neutrino-Nucleus Scattering (CEvNS) phenomenon.

The $\nu$-Angra experiment shares its experimental space, housed in an adapted container, with the CONNIE experiment \cite{Aguilar-Arevalo2019}. This collaborative setup constitutes a genuine laboratory dedicated to the study of neutrinos, optimizing the use of available resources and fostering a productive environment for neutrino research.

\section{Objectives}

The $\nu$-Angra experiment aims to develop a reliable and cost-effective technology for monitoring nuclear reactors and potentially the spectral evolution of neutrinos, which could reveal changes in fuel composition, particularly plutonium fractions, crucial for diversion detection. The specific objectives include:
\begin{itemize}
    \item Non-invasive monitoring of reactor activity.
    \item Estimation of the thermal power produced in the reactor core.
    \item Development of new antineutrino detection techniques.
    \item Contribution to the International Atomic Energy Agency (IAEA) safeguards and non-proliferation efforts.
    \item Integration of Latin American scientists and engineers into global scientific collaborations.
\end{itemize}

\section{Methodology}

The $\nu$-Angra experiment uses a water Cherenkov detector to monitor the electron antineutrino flux generated by a nuclear reactor via inverse beta decay (IBD) interactions $p + \overline{\nu}_e \rightarrow e^+ + n$. Positioned 25 meters away, a 1-ton detector records approximately 5000 events daily, aiding in antineutrino detection studies for safeguards. In the subsequent sections, we provide an overview of the detector components and their functional principles, along with an explanation of how event selection and data analysis are conducted.

\subsection{Detector Description}

The $\nu$-Angra detector comprises a 1-ton target volume filled with GdCl\(_3\)-doped water. The target volume is surrounded by 32 eight-inch photomultiplier tubes (PMTs) to detect Cherenkov radiation produced by IBD events. The detector also includes a Top VETO and a Lateral VETO system to reduce background noise from cosmic muons. A comprehensive description of the detector and readout electronics can be found in reference \cite{Lima2019}.

In the IBD interaction, an electron antineutrino ($\overline{\nu}_e$) interacts with a proton ($p$), producing a positron ($e^+$) and a neutron ($n$): $\nu_e + p \rightarrow e^+ + n$. The positron produces a Cherenkov flash and annihilates with an electron, creating two photons of approximately 511 keV each: $e^+ + e^- \rightarrow \gamma + \gamma$. This initial light burst is the 'prompt signal'. The neutron is captured by a gadolinium nucleus \cite{Hagiwara2019} within about 12.3 µs \cite{Lima2019}, emitting photons as the Gd atom deexcites. The reaction $n(\text{Gd, Gd}^*)\gamma$ typically releases around 8 MeV. This photon emission is the 'delay signal'. 

To reduce background noise, the detector's target is encased within two veto subsystems: the top veto above the target and the lateral veto around it. Both subsystems contain ultra-pure water and have four PMTs each. Simultaneous activation of two PMTs in either subsystem indicates background particle detection, causing the data acquisition system to stop recording. In addition, the lateral veto subsystem is encased in a non-active volume known as the 'shield', which features two sides measuring 14.5 cm thick and two others 22.5 cm thick, all filled with water. This configuration significantly improves the ability of the subsystem to block neutrons from cosmic radiation or the environment. The principal components of the $\nu$-Angra detector are illustrated in Figure 1.

\begin{figure}[h]
\centering
\includegraphics[width=0.6\textwidth]{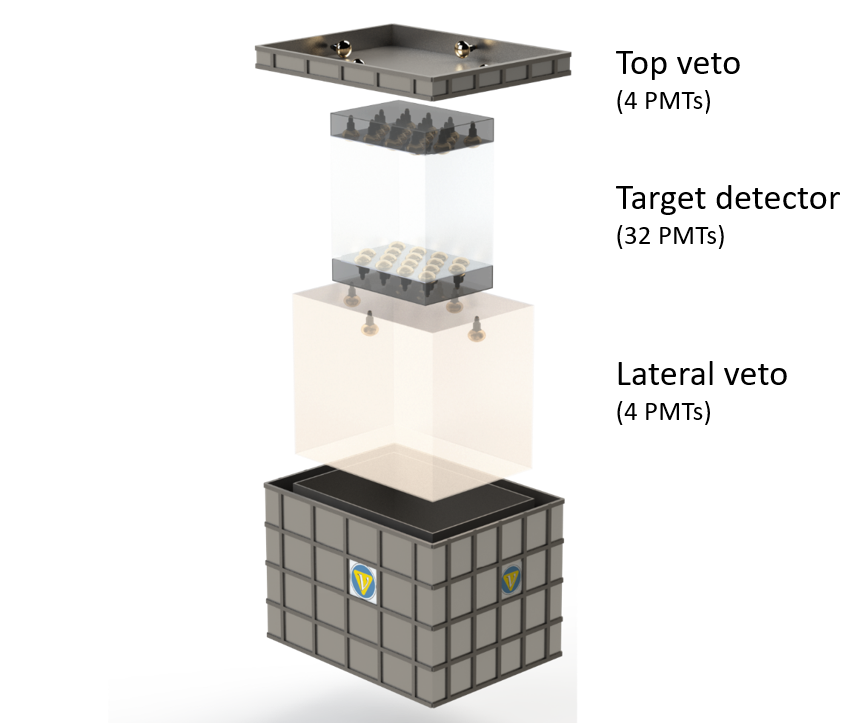}
\caption{Exploded view of the $\nu$-Angra detector showing the target volume and VETO systems.}
\label{fig:detector}
\end{figure}

\subsection{Readout Electronics}

\begin{figure}[!ht]
 \centering
 \begin{subfigure}{.35\textwidth}
   \centering 
   \includegraphics[width=\textwidth]{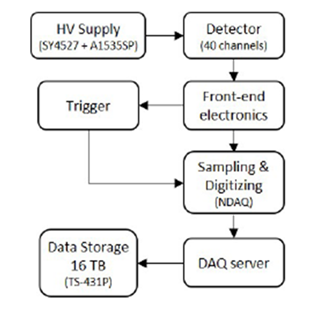}
   \caption{Electronics diagram}
   \label{fig:ElectronicDiagram} 
 \end{subfigure}
 \hfill
 \begin{subfigure}{.55\textwidth}
   \centering 
   \includegraphics[width=\textwidth]{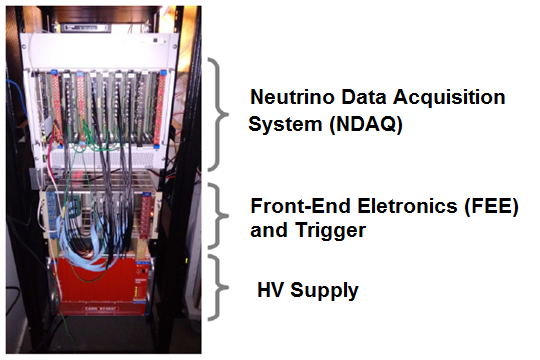}
   \caption{$\nu$-Angra Detector electronics rack}
   \label{fig:ElectronicRack} 
 \end{subfigure}
 \caption{Readout system diagram and a picture of the electronics rack installed in the neutrino laboratory at the Angra II nuclear power plant.}
 \label{fig:ReadoutSystem}
\end{figure}

The readout electronics consists of several key components: Front-End Electronics, digitizers, and the trigger system. The high voltage system and local computing infrastructure (DAQ servers and data storage) are commercial off-the-shelf components. The Front-End Electronics (FEE) \cite{FEE2016} and the digitizers (Neutrino Data Acquisition modules - NDAQ) \cite{NDAQ2014} were entirely designed and constructed by the collaboration, emphasizing the experiment's contribution to generating technology and scientific expertise.  The CAEN\footnote{\url{www.caen.it}} mainframe-based system provides high voltage for the 40 PMTs, controlled remotely via Ethernet. Data from the detector is continuously written to local storage and subsequently transferred to primary and mirror data servers located at CBPF (Rio de Janeiro) and Unicamp (Campinas), respectively. Figure \ref{fig:ElectronicRack} shows a picture of the rack with the $\nu$-Angra electronics installed in our laboratory at the reactor site.

\subsection{Calibration and Event Selection}

Due to restrictions on using radioactive sources at the reactor site, the calibration of the $\nu$-Angra detector is based on Monte Carlo simulations. This approach involves simulating the detector's response to antineutrino interactions, allowing for accurate energy calibration \cite{Santos2023}. Event selection is based on specific criteria to identify IBD events:
\begin{itemize}

    \item Energy of the \textit{prompt} and \textit{delay} events: The detector energy scale is calibrated using Monte Carlo simulations to accurately reflect the energy of the incoming antineutrinos, obtained from the simulated prompt event energy, obtained by positron simulations using as input the electron antineutrino spectrum as expected from the nuclear fuel \cite{Huber2004}, after the proper energy conversion given by $E_{e+} = E_{\Bar{\nu}} - \Delta ~MeV$, 
    where $\Delta = m_n - m_p = 1.293 ~MeV$, represents the mass difference between a neutron and a  proton, $E_e+$ and $E_{\Bar{\nu}}$ are the positron and the electron antineutrino energies respectively. The smeared line of the 8 MeV capture gammas from $n(Gd, Gd^*)\gamma$ is also identified by Monte Carlo calibration.
    
    \item Time correlation: The detector has sufficient temporal resolution and sensitivity to separate three distinct components: muon decays, neutron captures, and noise. This ability is crucial to accurately identify IBD events \cite{Lima2019}.
\end{itemize}

These criteria are used to select \textit{prompt-delay} pairs and ensure correct monitoring of reactor activity by comparing data collected with the reactor in the ON and OFF states.

\subsection{Data Analysis}

The data sets are previously submitted for reconstruction to extract from the raw data all the information required for analysis, such as timestamp, trigger multiplicity, and integrated charge from the PMTs. IBD event candidates are identified through the \textit{prompt-delay} pair selection as described in the previous section. The stability of the detector is checked by monitoring different quantities over time, such as trigger rate and parameters describing the shape of the reconstructed Michel electrons (ME) spectrum \cite{Lima2019}. The last is an unbiased check of the stability since is not related to the reactor's operational state. We verified the top veto tank's efficiency, averaging 99.76\% \cite{Souza2021}. The ME rate is 0.02 Hz, or 0.01\% of the total trigger rate, within the 0.24\% of unvetoed events. Not all unvetoed muons have enough energy to be stopped in the target, so these two numbers are consistent.

\subsubsection{ON-OFF Analysis}

\begin{figure}[!ht]
 \centering
 \begin{subfigure}{.45\textwidth}
   \centering 
   \includegraphics[width=\textwidth]{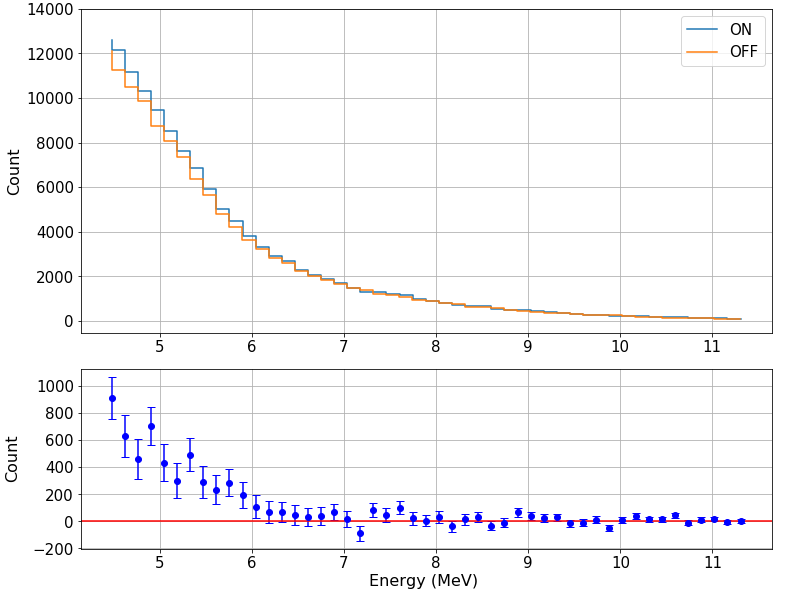}
   \caption{ON1-OFF analysis}
   \label{fig:ONOFF_Aug2020} 
 \end{subfigure}
 \hfill
 \begin{subfigure}{.47\textwidth}
   \centering 
   \includegraphics[width=\textwidth]{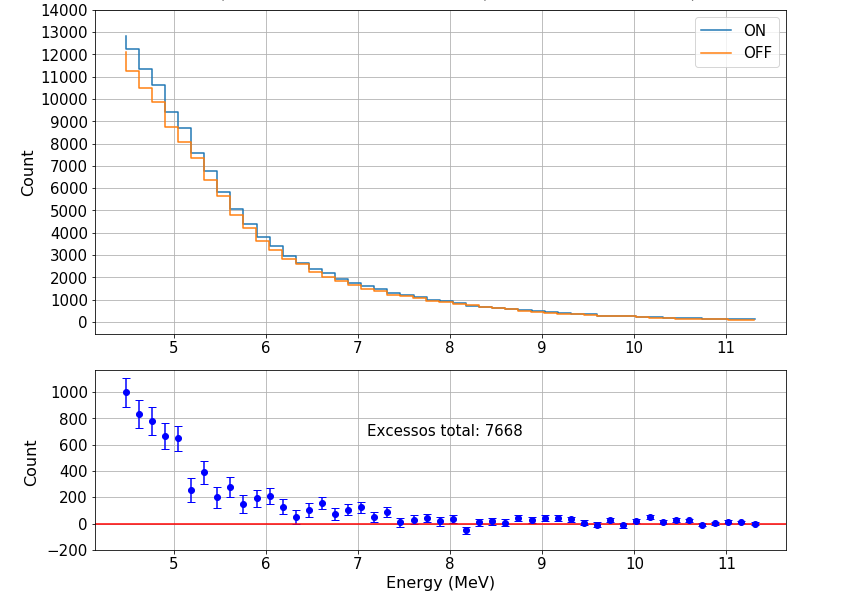}
   \caption{ON2-OFF analysis}
   \label{fig:ONOFF_Oct2020} 
 \end{subfigure}
 \caption{ON-OFF analysis results showing event rates comparison during two reactors ON and OFF periods (see text for details). The excess of events during the ON1 and ON2 periods is clear.}
 \label{fig:ONOFF}
\end{figure}

The ON-OFF analysis is critical to discern the presence of IBD events correlated with reactor activity. Data collected during the reactor's operational period (ON) is compared with data collected when the reactor is shut down (OFF). The \textit{prompt} spectra obtained in both cases are subtracted from each other (ON-OFF). We performed the analysis using two 30-day length data sets: August-September/2020 (ON1), September-October/2020 (ON2), and July-August/2020 (OFF). This method effectively highlights the excess of events during the ON period, as shown in Figure \ref{fig:ONOFF} confirming the detection of reactor antineutrinos.

\subsubsection{Results}

The findings of the ON-OFF analysis indicate a clear and statistically significant increase in events during the reactor ON period as opposed to the OFF period. An observable excess in the 4.5 - 6.5 MeV range aligns perfectly with the predicted energy spectrum of antineutrino events from IBD. A $\chi^2$ test was performed with H0 as the OFF spectrum, suggesting any excess could be due to statistical fluctuations. The null hypothesis, $H_0$, was unequivocally rejected for the ON1 period, as evidenced by a test statistic of $\chi^2/\text{dof} \sim$ 13.4, which corresponds to a nearly zero p-value. The same analysis was conducted for the ON2 period, as illustrated in Figure 19. Here, a test statistic of $\chi^2/\text{dof} \sim$ 17.3 continued to rebut H0, thereby validating the findings as a legitimate excess.

These excesses are attributed to the detection of reactor antineutrinos through the IBD process, validating the detector's performance and its potential for non-intrusive reactor monitoring.

\section{Current Status and Expected Challenges}

\subsection{Current status}
Due to the pandemic, maintenance visits to the detector became challenging, with most interventions carried out remotely via software. Despite UPS systems, some components of the trigger logic and FEE programming became unstable. Recently, data collection was interrupted due to insufficient human resources for operating the detector and conducting on-site maintenance. We observed a shortage of low-energy events and a shift to higher energy of the peak in the ME spectrum, suggesting the trigger favors more energetic events. The stable single photon electron (SPE) spectra of the PMTs imply a possible loss of water transparency rather than a deviation in PMT gains.

In July 2024, the detector was reactivated and readjusted to the operational parameters of October 2023 and has been collecting data steadily. We seem to have regained operational status but lack sufficient data to confirm a full return to proper condition. Minimal adjustments to the operating parameters are expected to maintain stable operation and robust data acquisition. We plan to sample and analyze the water to check the transparency loss.

\subsection{Expected challenges and upgrade perspectives}

The $\nu$-Angra experiment faces several challenges primarily due to its installation on the surface level, leading to high background noise from cosmic muons and other environmental sources. Despite these challenges, the experiment has demonstrated significant potential in reactor monitoring. To further enhance its capabilities, the following upgrades are proposed:

\subsubsection{Background Noise Reduction}

The current setup heavily relies on VETO systems to reduce background noise. An upgrade to improve the efficiency of these systems includes increasing the coverage and sensitivity of the VETO detectors. Implementing advanced algorithms for noise discrimination and enhancing the shielding materials around the detector can also contribute to better background rejection. One option is to install planes of segmented scintillators on the top and bottom of the detector to better differentiate muons and other background particles from the environment. Additionally, the reconstructed geometry of the muon tracks can be used to check detector stability, and a better understanding of the environment's background radiation will significantly improve the ability to distinguish the signal from the dominant background.

\subsubsection{Restoring stability and robust operation}

Restoring stable and robust operations is mandatory for two reasons:
\begin{enumerate}
    \item The detector needs optimal gain and stability to collect data before, during, and after the annual reactor shutdown for fuel exchange and maintenance, demonstrating the non-intrusive monitoring of the reactor's status.
    \item Recently, the power plant operator has initiated the transport of used nuclear fuel to a dedicated storage area. The reactor core elements that have been discarded are taken out of the reactor dome and handled in an area near the detector. The operations are conducted at distances of around 50m. At the beginning of such operations, when the detector was still running, we observed a large increase in the trigger rate and the events spectrum was compatible with the expected one for neutron captures.  This operation, supervised by both IAEA and ABACC, strictly adheres to safety regulations and ensures meticulous control of residual radiation levels. Our objective is to quantify the detector's heightened sensitivity to any observed increases in neutron capture rate. By doing so, we not only reinforce safety measures but also augment the array of safeguards with an enhanced capability stemming from the same monitoring device. Currently, we are performing dedicated measurements in coordination with the local Eletronuclear team.
\end{enumerate}

\subsubsection{Enhanced Calibration Methods}

The current calibration method relies on Monte Carlo simulations due to restrictions on using radioactive sources. Future upgrades could include the development of more sophisticated simulation models and the potential use of portable calibration sources that comply with safety regulations. Additionally, integrating in-situ calibration techniques using remote-controlled LEDs with different wavelengths could enhance the accuracy of the energy scale calibration.

\subsubsection{Improved Data Acquisition System}

The data acquisition (DAQ) system and the front-end electronics (FEE) have been custom-designed by the collaboration. However, upgrading the DAQ system to incorporate faster and more efficient data processing capabilities, such as FPGA-based trigger systems and real-time data analysis frameworks, can significantly improve the performance and reliability of the experiment.
Upgrading the instrumentation electronics for the Neutrinos-Angra experiment involves complex challenges due to rare neutrino events and the need for efficient detection. Detection relies on identifying specific signatures in the signal from the detector. Current front-end electronics have limitations in increasing event acquisition rates, crucial for better detection efficiency but entails added complexity and cost. Our goal is to train a neural network to create an AI model for real-time event identification. The challenge is implementing these AI models on the front-end electronics' FPGA. This aims to enhance the detector's trigger system efficiency for effective neutrino detection, as seen in the Angra-Neutrinos experiment. Enhanced efficiency is vital for both fundamental science and practical applications, like monitoring radioactive materials, crucial to the IAEA and global security. Successful technology implementation will improve nuclear fuel safety and control in nuclear plants and submarines, advancing science and global security. Thus, AI integration in data acquisition electronics is essential for next-generation neutrino detectors, ensuring superior sensitivity and precision in critical contexts.

\subsubsection{Replacing Water with WBLS}

Replacing the current water-based Cherenkov detector with a Water-based Liquid Scintillator (WBLS) \cite{Xiang:2024} offers a promising upgrade to enhance the energy resolution of the $\nu$-Angra detector. WBLS combines the benefits of both water Cherenkov detectors and liquid scintillators, providing improved light yield and timing characteristics while maintaining a high level of transparency.

\textbf{Improved Energy Resolution:} The enhanced light yield from WBLS allows for better energy resolution, which is crucial for accurate measurement of the energy spectrum of detected antineutrinos. This improvement enables the detector to more precisely determine the energy of each event, leading to better discrimination between signal and background.

\textbf{Fuel Content Measurement:} With improved energy resolution, the $\nu$-Angra detector can potentially measure the evolution of the reactor's fuel composition. By following the spectral evolution of the electron antineutrinos, it is possible to infer changes in the fuel content, such as the burn-up of uranium and the build-up of plutonium \cite{Huber2004}. This capability is significant for non-proliferation efforts, as it provides a non-intrusive method to monitor the reactor's fuel cycle and detect any anomalies that may indicate unauthorized activities.

\subsection{The Cryogenic Calorimeter}

One of the proposed future enhancements is the installation of a cryogenic calorimeter to detect neutrinos via Coherent Elastic Neutrino-Nucleus Scattering (CEvNS), replacing the water-based detector. This new device would not only improve the sensitivity of the detector, but also expand its research capabilities to explore new physics phenomena.

The utilization of milliKelvin (mK) calorimeters for the detection of rare event decays, originally proposed by Fiorini and Niinikoski in 1984 \cite{Fiorini:1983yj}, has undergone substantial advancements. This highly sensitive and effective methodology has been used in large-scale experiments, such as CUORE/CUPID \cite{CUORE:2021mvw, CUPID:2022jlk} for neutrinoless double-beta decay and CRESST \cite{CRESST:2023cgp} for sub-GeV dark matter detection.

Cryogenic calorimeters work by converting particle energy in a crystal into phonons, which are collected by a thermometer to measure a temperature rise. This allows the use of different isotopes with different crystal materials, making the technology versatile.

\subsubsection{Application in the $\nu$-Angra Experiment}
Replacing the current water-based Cherenkov detector with a cryogenic calorimeter in the next five years presents a strategic enhancement for the $\nu$-Angra experiment. The integration of this advanced technology can significantly improve the experiment's capabilities:

\textbf{Sensitivity and Flexibility: } Cryogenic calorimeters can achieve ultra-low thresholds, with macroscopic detectors capable of detecting energy deposits below 10 eV. This high sensitivity makes them ideal for detecting low-energy events such as Coherent Elastic Neutrino-Nucleus Scattering (CEvNS).

\textbf{Modularity and Scalability: } The technology allows for building large detectors with significant exposure while maintaining low thresholds, making it scalable for extensive experimental setups.

\subsubsection{Fundamental Physics Research}
Cryogenic calorimeters offer great potential to advance fundamental physics research. By using these detectors to study CEvNS, we can investigate the theory behind these interactions. This is possible because we can change the detector target (the crystal), and the cross-section depends strongly on the material's mass number $A$. Such studies can provide deeper insights into fundamental forces and particle interactions.

\subsubsection{Required Infrastructure}

To develop and implement such a detector at the Angra dos Reis reactor, a relatively compact infrastructure is required:

\begin{itemize}
    \item \textbf{Cryostat: } A high-end dry-cryostat with noise reduction capabilities is essential. This could be provided by manufacturers like Leiden Cryogenics\footnote{\url{https://leidencryogenics.nl}}.
    \item \textbf{Veto System and DAQ: } An effective veto system and a data acquisition system (DAQ) to record signals are necessary. The expertise of the Cherenkov detector operations ensures the quality, efficacy, and reliability of these components.
    \item \textbf{Space Requirements: } The entire setup can fit inside a high cube container, making it possible to install it at the reactor site.
\end{itemize}

\subsection{Integration with Global Networks}

Enhancing the integration of the $\nu$-Angra detector with global neutrino monitoring networks can provide a more comprehensive understanding of neutrino behavior and reactor operations. This integration could include real-time data sharing and collaborative analysis efforts with other international experiments.


\section{Timeline}

This timeline outlines a five-year scenario in which the existing Cherenkov detector continues to operate and collect data while the development and testing of a cryogenic calorimeter take place in parallel. After five years, the cryogenic calorimeter system will replace the Cherenkov detector, allowing for a seamless transition and significant enhancement of the $\nu$-Angra experiment's capabilities.

\subsection{Year 1-5: Operation and Upgrades of Cherenkov Detector}
\begin{itemize}
    \item \textbf{Operation and Data Collection (Years 1-5):}
    \begin{itemize}
        \item Continue operation of the current water-based Cherenkov detector for reactor monitoring and data collection.
        \item Perform regular maintenance and calibration to ensure optimal performance.
    \end{itemize}
    \item \textbf{System Enhancements (Years 2-5):}
    \begin{itemize}
        \item Implement system upgrades to improve sensitivity and data quality.
        \item Introduce software and hardware improvements to improve data acquisition and processing.
    \end{itemize}
\end{itemize}

\subsection{Year 1-3: Planning and Initial Development of Cryogenic Calorimeter}
\begin{itemize}
    \item \textbf{Feasibility Study and Proposal Development (Year 1):}
    \begin{itemize}
        \item Conduct a detailed feasibility study to assess the potential and requirements for replacing the Cherenkov detector with a cryogenic calorimeter.
        \item Develop a comprehensive project proposal and secure funding.
    \end{itemize}
    \item \textbf{Design and Specification (Years 2-3):}
    \begin{itemize}
        \item Develop detailed designs for the cryogenic calorimeter system.
        \item Specify the requirements for the cryostat, detectors, DAQ system, and veto system.
    \end{itemize}
\end{itemize}

\subsection{Year 3-5: Development and Testing of Cryogenic Calorimeter}
\begin{itemize}
    \item \textbf{Component Procurement and Prototype Development (Years 3-4):}
    \begin{itemize}
        \item Procure necessary components and begin the development of prototype systems.
        \item Conduct initial testing and optimization of the prototypes.
    \end{itemize}
    \item \textbf{System Integration and Preliminary Testing (Year 5):}
    \begin{itemize}
        \item Integrate all components into a working prototype system.
        \item Perform preliminary testing to ensure system functionality and performance.
    \end{itemize}
\end{itemize}

\subsection{Post Year 5: Assembly and Commissioning of Cryogenic Calorimeter}
\begin{itemize}
    \item \textbf{Site Preparation and Installation (Post Year 5):}
    \begin{itemize}
        \item Prepare the installation site at the Angra dos Reis reactor.
        \item Install the cryogenic calorimeter system and perform initial setup checks.
    \end{itemize}
    \item \textbf{Calibration and Full Commissioning (Post Year 5):}
    \begin{itemize}
        \item Calibrate the system using known sources and Monte Carlo simulations.
        \item Complete full commissioning and transition to regular data collection.
    \end{itemize}
\end{itemize}


\section{Costs}

\subsection{Upgrades and Running Costs of Cherenkov Detector}

In parallel with the development of the cryogenic calorimeter, the Cherenkov detector will continue to operate and undergo necessary upgrades. The following costs are associated with these activities:

\begin{itemize}
    \item \textbf{System Upgrades:} Regular upgrades to improve sensitivity, data quality, and hardware performance. Estimated cost: \$180,000.
    \item \textbf{Maintenance and Calibration:} Routine maintenance and calibration to ensure optimal performance. Estimated cost: \$60,000.
    \item \textbf{Operational Costs:} Running costs including power, consumables, and staff. Estimated annual cost: \$30,000.
\end{itemize}

\textbf{Subtotal Cherenkov Detector:} \$270,000

These investments will ensure continued operation and enhancement of the Cherenkov detector, providing high-quality data while the cryogenic calorimeter system is being developed and tested.

\subsection{Implementation of Cryogenic Calorimeter System}

The implementation of the cryogenic calorimeter system involves several cost components which are outlined below.

\begin{itemize}
    \item \textbf{Cryostat:} A high-end dry-cryostat with noise reduction capabilities is essential for the operation of the cryogenic calorimeter. The estimated cost for the cryostat is approximately \$450,000.
    \item \textbf{Detectors and DAQ System:} The detectors and the data acquisition (DAQ) system are crucial for capturing and processing the signals from the calorimeter. The combined cost for the detectors and DAQ system is estimated to be less than \$55,000.
    \item \textbf{Veto System:} An effective veto system is necessary to reduce background noise and enhance the sensitivity of the calorimeter. The cost of the veto system is around \$110,000.
    \item \textbf{Infrastructure:} The necessary infrastructure to house the cryogenic calorimeter system, including site preparation and support structures, is estimated to cost around \$55,000.
\end{itemize}

\textbf{Subtotal Cryogenic Calorimeter System:} \$670,000

\subsection{Total Estimated Costs}

Combining the costs for ongoing upgrades and running costs of the Cherenkov detector with the costs for the cryogenic calorimeter system, the total estimated budget for the next five years is outlined below.

\begin{itemize}
    \item \textbf{Cherenkov Detector Upgrades and Running Costs:} 
    \begin{itemize}
        \item System Upgrades: \$180,000
        \item Maintenance and Calibration: \$60,000
        \item Operational Costs: \$30,000
        \item \textbf{Subtotal Cherenkov Detector:} \$270,000
    \end{itemize}
    \item \textbf{Cryogenic Calorimeter System:} 
    \begin{itemize}
        \item Cryostat: \$450,000
        \item Detectors and DAQ System: \$55,000
        \item Veto System: \$110,000
        \item Infrastructure: \$55,000
        \item \textbf{Subtotal Cryogenic Calorimeter System:} \$670,000
    \end{itemize}
\end{itemize}

\textbf{Total Estimated Costs:} \$940,000

These estimates are based on current market prices and expected expenses. Although they are realistic, the final costs may vary due to unforeseen circumstances, market fluctuations, or additional requirements that may arise during the project.


\section{Computational Requirements}

The computational infrastructure of the institutions involved in the $\nu$-Angra experiment, mainly CBPF, UNICAMP, and USP in the upcoming future, is excellent and fully capable of meeting the needs of the experiment. These institutions have robust computational resources that support the acquisition, processing, and analysis of data required for the project.

The existing resources currently have enough capacity to continue supporting the experiment. However, to ensure the longevity and reliability of the computational infrastructure, it is advisable to perform upgrades on the data acquisition machines, connection servers, software servers, and storage servers. These upgrades will improve performance, security, and data handling capabilities.

The anticipated cost for the necessary hardware upgrades and maintenance of the computational infrastructure is estimated to be around \$50,000. This investment will help maintain the high standards of data integrity and processing efficiency required for the successful operation of the $\nu$-Angra experiment.


\section{Conclusion}

The $\nu$-Angra experiment, to our knowledge, was a pioneer in the detection of reactor neutrinos using a Cherenkov detector that runs at the surface level, proving the efficiency of neutrino-based safeguards.  Upgrades to the Cherenkov detector and the development of a cryogenic calorimeter highlight our commitment to enhancing detection and expanding research scopes. These technologies will strengthen non-proliferation efforts and provide insights into neutrino physics, establishing the $\nu$-Angra experiment as a key player in the scientific community.

Furthermore, the technologies to be developed for the cryogenic detector, based on quantum sensors, hold immense potential for applications in industrial X-ray detectors, the aerospace sector, and the development of qubits. These advances will have far-reaching impacts beyond the immediate scope of neutrino detection, driving innovation across multiple high-tech industries.

 The $\nu$-Angra experiment has also generated a significant number of master's and doctoral theses, underscoring its importance in training human resources in advanced technologies, both in hardware and software. This contribution highlights the role of the experiment in the promotion of the next generation of scientists and engineers equipped with cutting-edge skills and knowledge.

The $\nu$-Angra experiment, alongside CONNIE, demonstrates the successful partnership with Eletronuclear at Angra II. This collaboration highlights Brazil and Latin America's capacity for advanced particle physics research without large-scale infrastructure investments. Using existing facilities, these experiments yield high-quality scientific results, showcasing the region's potential for world-class research.

These efforts collectively underscore the transformative potential of the $\nu$ -Angra experiment and, in a broader context, the neutrino laboratory at the Angra II power plant, signaling a new era of precision and innovation in neutrino research.


\end{document}